\documentclass[11pt]{article}

\usepackage{epsfig,latexsym}
\usepackage{amsmath, amssymb, amsthm}
\usepackage{tkz-graph}
\usepackage{tikz}
\usetikzlibrary{decorations.pathreplacing}

\newtheorem{theorem}{Theorem}
\newtheorem{lemma}[theorem]{Lemma}
\newtheorem{definition}{Definition}

\newtheorem{proposition}{Proposition}
\newtheorem{problem}{Open Problem}
\newtheorem{conjecture}{Conjecture}
\newtheorem{corollary}{Corollary}

\title{More results on weighted independent domination\footnote{Extended abstract of this paper appeared in the proceedings of 
WG 2017 -- the 43rd International Workshop on Graph-Theoretic Concepts in Computer Science \cite{WG2017}.}}

\author{Vadim Lozin\thanks{Mathematics Institute, University of Warwick, Coventry CV4 7AL, UK. E-mail: V.Lozin@warwick.ac.uk.}
\and
Dmitriy Malyshev\thanks{National Research University Higher School of Economics, 25/12 Bolshaya Pecherskaya Ulitsa, 603155, Nizhny Novgorod, Russia.
E-mail: dmalishev@hse.ru.}
\and
Raffaele Mosca\thanks{Dipartimento di Economia, Universit\'{a} degli Studi ``G.~D'Annunzio'', Pescara 65121, Italy.
E-mail: R.Mosca@unich.it.}
\and 
Viktor Zamaraev\thanks{Mathematics Institute, University of Warwick, Coventry CV4 7AL, UK.
E-mail: V.Zamaraev@warwick.ac.uk.}
}

\tikzstyle{b_vertex}=[circle,fill=black!100,text=white,inner sep=0.8mm,draw]
\tikzstyle{g_vertex}=[circle,fill=black!10,text=black,inner sep=0.8mm,draw]
\tikzstyle{w_vertex}=[circle,fill=white!100,text=black,inner sep=0.8mm,draw]
\tikzstyle{point}=[circle,fill=black,inner sep=0.1mm]
\tikzstyle{path_edge}=[thick]

\newtheorem{observation}{Observation}

\def\forbGraphs{$(P_5, \overline{P_3+P_2)}$}
\def\compPfiveStar{\overline{P_5}^*}
\def\threeForbGraphs{$(P_5, \overline{P_3+P_2}, \compPfiveStar)$}
\def\compPthreePtwo{\overline{P_3+P_2}}

\begin{document}

\date{}
\maketitle

\begin{abstract}
Weighted independent domination is an NP-hard graph problem, which remains computationally intractable 
in many restricted graph classes. In particular, the problem is NP-hard in the classes of sat-graphs
and chordal graphs. We strengthen these results by showing that the problem is NP-hard in a proper subclass of 
the intersection of sat-graphs and chordal graphs. On the other hand, we identify two new classes of graphs 
where the problem admits polynomial-time solutions. 
\end{abstract}

\section{Introduction}

\textsc{Independent domination} is the problem of finding in a graph an inclusionwise maximal independent set of minimum cardinality. 
This is one of the hardest problems of combinatorial optimization and it remains difficult under substantial restrictions. 
In particular, it is NP-hard for so-called sat-graphs, where the problem is equivalent to {\sc satisfiability} \cite{Zverovich06}. 
It is also NP-hard for planar graphs, triangle-free graphs, graphs of vertex degree at most 3 \cite{BL03}, line graphs \cite{YG80},  chordal bipartite graphs \cite{DMK90}, etc.

The weighted version of the problem (abbreviated WID) deals with vertex-weighted graphs and asks to find  an inclusionwise maximal independent set of minimum total weight.
This version is provenly harder, as it remains NP-hard even for chordal graphs \cite{Chang2004}, where {\sc independent domination} can be solved in polynomial time \cite{Farber}. 
In the present paper, we strengthen two NP-hardness results by showing that WID is NP-hard in a proper subclass of the intersection of sat-graphs and chordal graphs.

On the positive side, it is known that the problem is polynomial-time solvable for interval graphs, permutation graphs \cite{poly}, graphs of bounded clique-width \cite{CW}, etc.

Let us observe that all classes mention above are hereditary,  i.e. closed under taking induced subgraphs. It is well-known 
(and not difficult to see) that a class of graphs is hereditary if and only if it can be characterized in terms of minimal forbidden induced subgraphs.
Unfortunately, not much is known about efficient solutions for the WID problem on graph classes defined by {\it finitely many} forbidden induced subgraphs. 
Among rare examples of this type, let us mention cographs and split graphs. 
\begin{itemize}
\item A {\it cograph} is a graph in which every induced subgraph with at least two vertices is either disconnected or the complement of a disconnected 
graph. The cographs are precisely $P_4$-free graphs, i.e. graphs containing no induced $P_4$. In the case of cographs, the problem can be solved efficiently by means of modular decomposition. 
\item A {\it split graph} is a graph whose vertices can be partitioned into a clique and an independent set. In terms of forbidden induced subgraphs, 
the split graphs are the graphs which are free of $2K_2, C_{4}$ and $C_5$.
The only available way to solve WID efficiently for a split graph is to examine all its inclusionwise maximal independent sets, of which there are polynomially many.
\end{itemize}

The class of sat-graphs, mentioned earlier, consists of graphs whose vertices can be partitioned into 
a clique and a graph of vertex degree at most 1. Therefore, sat-graphs form an extension of split graphs. With this extension the complexity status of the problem
jumps from polynomial-time solvability to NP-hardness. In the present paper, we study two other extensions of split graphs and show polynomial-time solvability in both of them.

The first of them deals with the class of $(P_5,\overline{P}_5)$-free graphs, which also extends the cographs. From an algorithmic point of view, 
this extension is resistant to any available technique. To crack the puzzle for $(P_5,\overline{P}_5)$-free graphs, we develop a new decomposition 
scheme combining several algorithmic tools. This enables us to show that the WID problem can be solved  
for $(P_5,\overline{P}_5)$-free graphs in polynomial time. 

The second extension of split graphs studied in this paper deals with the class of \forbGraphs-free graphs. 
To solve the problem in this case, we develop a tricky reduction allowing us to reduce the problem to the first class.

Let us emphasize that in both cases the presence of $P_5$ among the forbidden graphs is necessary, 
because each of $\overline{P}_5$ and $\compPthreePtwo$ contains a $C_4$ and by forbidding $C_4$ alone we obtain a class where the problem is NP-hard \cite{BL03}.
Whether the presence of $P_5$ among the forbidden graphs is sufficient for polynomial-time solvability of WID is a big open question. 
For the related problem of finding a maximum weight independent set (WIS), this question was answered only recently \cite{P5} after several decades of attacking 
the problem on subclasses of $P_5$-free graphs (see e.g. \cite{gem,PP,Kar}). In particular, prior to solving the problem for $P_5$-free graphs, it was solved for 
$(P_5,H)$-free graphs for all graphs $H$ with at most 5 vertices, except for $H=C_5$.

WID is a more stubborn problem, as it remains NP-hard in many classes where WIS can be solved in polynomial time, such as line graphs, chordal graphs, bipartite graphs, etc.
In \cite{LozMosPur2015}, the problem was solved in polynomial time for many subclasses of $P_5$-free graphs, including $(P_5,H)$-free graphs for all graphs $H$ with at most 5 vertices, 
except for $H=\overline{P}_5$, $H=\compPthreePtwo$ and $H=C_5$. In the present paper, we solve the first two of them, leaving the case of $(P_5,C_5)$-free graphs open.
We believe that WID in $(P_5,C_5)$-free graphs is polynomially equivalent to WID in $P_5$-free graphs. Determining the complexity status of the problem in both classes is a challenging open question. 
We discuss this and related open questions in the concluding section of the paper.  

The rest of the paper is organized as follows. In the remainder of the present section, we introduce basic terminology and notation. 
In Section~\ref{sec:house} we solve the problem for $(P_5,\overline{P}_5)$-free graphs,
and in Section~\ref{sec:new} we solve it  for \forbGraphs-free graphs.


\medskip
All graphs in this paper are finite, undirected, without loops and multiple edges. 
The vertex set and the edge set of a graph $G$ are denoted by $V(G)$ and $E(G)$, respectively.
A subset $S\subseteq V(G)$ is 
\begin{itemize}
\item[--] \textit{independent} if no two vertices of $S$ are adjacent,
\item[--] a \textit{clique} if every two vertices of $S$ are adjacent,
\item[--] \textit{dominating} if every vertex not in $S$ is adjacent to a vertex in $S$.
\end{itemize}

For a vertex-weighted graph $G$ with a weight function $w$, by $id_w(G)$ we denote the minimum weight of an independent dominating set in $G$.

If $v$ is a vertex of $G$, then $N(v)$ is the {\it neighbourhood} of $v$ (i.e. the set of vertices adjacent to $v$)
and $V(G) \setminus N(v)$ is the {\it antineighbourhood} of $v$. 
We say that $v$ is  \textit{simplicial} if its neighbourhood is a clique, and $v$ is \textit{antisimplicial} if 
its antineighbourhood is an independent set.

Let $S$ be a subset of $V(G)$. We say that a vertex $v \in V(G) \setminus S$ \textit{dominates} $S$ if $S\subseteq N(v)$.
Also, $v$ \textit{distinguishes} $S$ if $v$ has both a neighbour and a non-neighbour in $S$.
By $G[S]$ we denote the subgraph of $G$ induced by $S$ and by $G - S$ the subgraph $G[V \setminus S]$.
If $S$ consists of a single element, say $S = \{ v \}$, we write $G - v$, omitting the brackets.

If $G$ is a connected graph but $G-S$ is not, then $S$ is a \textit{separator} (also known as a cut-set). 
A \textit{clique separator} is a separator which is also a clique. 

As usual, $P_n,C_n$ and $K_n$ denote a chordless path, a chordless cycle and a complete graph on $n$ vertices, respectively. 
Given two graphs $G$ and $H$, we denote by $G+H$ the disjoint union of $G$ and $H$, and by $mG$ 
the disjoint union of $m$ copies of $G$. 

We say that a graph $G$ contains a graph $H$ as an induced subgraph if $H$ is isomorphic to an induced subgraph of $G$.
Otherwise, $G$ is $H$-free.

A class $\mathcal{Z}$ of graphs is hereditary if it is closed under taking induced subgraphs, i.e. if 
$G \in \mathcal{Z}$ implies that every induced subgraph of $G$ belongs to $\mathcal{Z}$. 
It is well-known that $\mathcal{Z}$ is hereditary if and only if graphs in $G$ do not contain induced subgraphs from a set $M$,
in which case we say that $M$ is the set of forbidden induced subgraphs for $\mathcal{Z}$.

For an initial segment of natural numbers $\{ 1, 2, \ldots, n \}$ we will often use the notation $[n]$.

\section{An NP-hardness result}
\label{sec:NP}
As we mentioned in the introduction, the WID problem is NP-hard in the classes of sat-graphs and chordal graphs. 
A graph is {\it chordal} if it is $(C_4,C_5,C_6,\ldots)$-free. 
A graph $G$ is called a \textit{sat-graph} if there exists a partition $A \cup B = V(G)$ such
that
\begin{enumerate}
	\item $A$ is a clique (possibly, $A = \emptyset$);
	\item $G[B]$ is an induced matching, i.e. an induced 1-regular graph (possibly, $B = \emptyset$);
	\item there are no triangles $(a,b,b')$, where $a \in A$ and $b,b' \in B$.
\end{enumerate}
We shall refer to the pair $(A,B)$ as a \textit{sat-partition} of $G$.

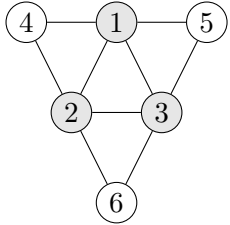
\begin{figure}[ht] 
     	\centering
     	\begin{tikzpicture}
			[scale=.6,auto=left]
	  		
%
%

	  		\node[g_vertex] (1) at (10,1) {$1$};
	  		\node[g_vertex] (2) at (9,-1) {$2$};
	  		\node[g_vertex] (3) at (11,-1) {$3$};
	  		\node[w_vertex] (4) at (8,1) {$4$};
	  		\node[w_vertex] (5) at (12,1) {$5$};
 	  		\node[w_vertex] (6) at (10,-3) {$6$};
 
			\foreach \from/\to in {1/2,1/4,1/3,1/5,2/4,2/6,2/3,3/5,3/6}
	    		\draw (\from) -- (\to);	
	    		
	\end{tikzpicture}
	    		
	\caption{Graph $Sun_3$} 
	\label{fig:T_domino}
\end{figure}

Below we show that WID is NP-hard in the class of $(C_4, Sun_3)$-free sat-graphs, where $Sun_3$ is the graph shown in Figure~\ref{fig:T_domino}. 
Since cycles $C_k$ with $k\ge 5$ are not sat-graphs (which is easy to see), this class also is a subclass of chordal graphs. Moreover, $Sun_3$ is 
both a sat-graph and a chordal graph. Therefore, $(C_4, Sun_3)$-free sat-graphs form a proper subclass of the intersection of sat-graphs and chordal graphs.


Before we prove the main result of this section, let us make the following useful observation.

\begin{observation}\label{obs:domino_T}
	Let $G$ be a sat-graph with a sat-partition $(A,B)$. If $G$ contains $Sun_3$ 
		as an induced subgraph, then $1,2,3 \in A$ and $4,5,6 \in B$.
\end{observation}


\begin{theorem}
	The WID problem is NP-hard in the class of $(C_4, Sun_3)$-free sat-graphs.
\end{theorem}
\begin{proof}
We prove the theorem by transforming the decision version of the {\sc minimum dominating set} problem 
in $(C_3,C_4,C_5,C_6)$-free graphs to the WID problem in $(C_4, Sun_3)$-free graphs.
Since the former problem in NP-complete (see \cite{Kor90}), this will prove that the latter is
NP-hard.

For an $n$-vertex graph $G = (V,E)$ let us define the graph $G' = (V',E')$ with vertex set
$V' = \{v_1,v_2,v_3 : v \in V\}$
and edge set

$E' = \{ (v_1,v_2), (v_2,v_3): v \in V\} 
\cup \{ (w_2,v_3), (w_3,v_2) : (w,v) \in E\} \cup \{(w_3, v_3) : w, v \in V, u \neq v\}$.

\begin{figure}[ht] 
     	\centering
     	\begin{tikzpicture}
			[scale=.6,auto=left]
	  		
	  		\node[w_vertex] (a) at (0,2) {$a$};
	  		\node[w_vertex] (b) at (5,2) {$b$};
	  		\node[w_vertex] (c) at (10,2) {$c$};
 	  		\node[w_vertex] (d) at (15,2) {$d$};
 
			\foreach \from/\to in {a/b, b/c, c/d}
	    		\draw (\from) -- (\to);	
	    		
	    		\coordinate [label=center:$P_4$] (P4) at (7.5,0.5);

	  		\node[w_vertex] (a3) at (0,-4) {\footnotesize{$a_3$}};
	  		\node[w_vertex] (b3) at (5,-4) {\footnotesize{$b_3$}};
	  		\node[w_vertex] (c3) at (10,-4) {\footnotesize{$c_3$}};
 	  		\node[w_vertex] (d3) at (15,-4) {\footnotesize{$d_3$}};

 	  		\node[w_vertex] (a2) at (0,-7) {\footnotesize{$a_2$}}; 	  		
	  		\node[w_vertex] (a1) at (0,-10) {\footnotesize{$a_1$}};

	  		\node[w_vertex] (b2) at (5,-7) {\footnotesize{$b_2$}};	  		
	  		\node[w_vertex] (b1) at (5,-10) {\footnotesize{$b_1$}};

	  		\node[w_vertex] (c2) at (10,-7) {\footnotesize{$c_2$}};	  		
	  		\node[w_vertex] (c1) at (10,-10) {\footnotesize{$c_1$}};
 
 	  		\node[w_vertex] (d2) at (15,-7) {\footnotesize{$d_2$}};
 	  		\node[w_vertex] (d1) at (15,-10) {\footnotesize{$d_1$}};
 
			\foreach \from/\to in {a3/b3,b3/c3,c3/d3,
								a3/a2,a2/a1,
								b3/b2,b2/b1,
								c3/c2,c2/c1,
								d3/d2,d2/d1,
								a3/b2,b3/a2,
								b3/c2,c3/b2,
								c3/d2,d3/c2}
	    		\draw (\from) -- (\to);
	    		
	    		\draw (a3) to[out=20, in=160] (c3);
	    		\draw (b3) to[out=20, in=160] (d3);
	    		\draw (a3) to[out=25, in=155] (d3);
	    		
	    		\coordinate [label=center:$P'_4$] (transP4) at (7.5,-11.5);
	\end{tikzpicture}
	    		
	\caption{Graphs $P_4$ (top) and $P'_4$ (bottom)} 
	\label{fig:transformationP4}
\end{figure}
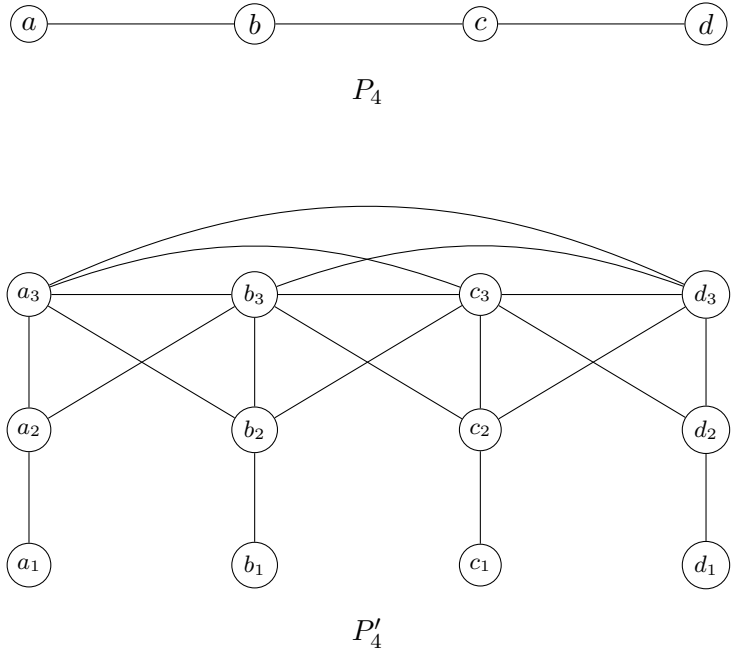

Figure~\ref{fig:transformationP4} illustrates the transformation of $P_4$ into $P'_4$.
It is easy to see that for every graph $G$, the graph $G'$ is a sat-graph. Moreover, it is $C_4$-free, i.e. $G'$ is a chordal graph.
Also using the fact that $Sun_3$ has the unique sat-partition (see Observation \ref{obs:domino_T}) it is not hard to check that if $G'$ 
contains $Sun_3$ as an induced subgraph, then $G$ has a cycle of length at most 6. Therefore, for any
$(C_3,C_4,C_5,C_6)$-free graph $G$, the graph $G'$ is a $(C_4, Sun_3)$-free sat-graph.

Further, for every $v \in V$ we assign weight 1 to vertex $v_1$, weight $2$ to vertex $v_2$, 
and weight $2n$ to vertex $v_3$.

Now, we claim that $G$ has a dominating set of size at most $k$ if and only if $G'$ has an independent dominating set of total weight at most $n + k$. 
First, suppose $G$ has a dominating set $D$ of size at most $k$.
Then $D' = \{v_2 : v \in D\} \cup \{v_1 : v \in V \setminus D\}$ is clearly an independent dominating set
of $G'$ with total weight at most $n + k$.
On the other hand, suppose $G'$ has an independent dominating set $D'$ of total weight at most 
$n + k$.
If $k \geq n$, then $V$ is a dominating set of $G$ of size at most $k$.
If $k < n$, then $D'$ cannot contain any of the vertices of weight $2n$ and hence $D'$ is of the form $\{v_2 : v \in D\} \cup \{v_1 : v \in V \setminus D\}$ for some subset $D$ of $V$.
For any vertex $u \in V$, since $u_3$ is dominated in $G'$ by some $v_2 \in D'$, we have that in $G$ vertex $u$ is dominated by $v \in D$.
Hence, $D$ is a dominating set of $G$. Moreover, the total weight of $D'$ is $n + |D|$ implying that $D$ is of size at most $k$.
\end{proof}

\section{WID in $(P_5,\overline{P}_5)$-free graphs}
\label{sec:house}
To solve the problem for $(P_5,\overline{P}_5)$-free graphs, we first develop a new decomposition scheme in Section~\ref{sec:had}
that combines modular decomposition (Section~\ref{subsec:modular}) and antineighborhood decomposition (Section~\ref{subsec:anti}). 
Then in Section~\ref{sec:P5} we apply it to  $(P_5,\overline{P}_5)$-free graphs.

\subsection{Graph decompositions}

\subsubsection{Modular decomposition}
\label{subsec:modular}

Let $G=(V,E)$ be a graph. A set $M \subseteq V$ is a $module$ in $G$ if no vertex outside of $M$ distinguishes $M$.
Obviously, $V(G)$, $\emptyset$ and any vertex of $G$ are modules and we call them {\it trivial}.  
A non-trivial module is also known as a \textit{homogeneous set}. A graph without homogeneous sets is called {\it prime}. 
The notion of a prime graph plays a crucial role in {\em modular decomposition},
which allows to reduce various algorithmic and combinatorial  problems in a hereditary class $\mathcal{Z}$ to prime graphs in $\mathcal{Z}$
(see e.g. \cite{MoeRad1984/1} for more details on modular decomposition and its applications).
In particular, it was shown in \cite{BL03} that the WID problem can be solved in polynomial time in 
$\mathcal{Z}$ whenever it is polynomially solvable for prime graphs in $\mathcal{Z}$.

In our solution, we will use homogeneous sets in order to reduce the problem from a graph $G$ to two proper induced subgraphs of $G$ as follows.
Let $M \subset V$ be a homogeneous set in $G$. Denote by $H$ the graph obtained from $G$ by contracting $M$ into a single vertex $m$ (or equivalently,
by removing all but one vertex $m$ from $M$). We define the weight function $w'$ on the vertices of $H$ as follows: 
$w'(v) = w(v)$ for every $v \ne m$, and $w'(m) = id_w(G[M])$. Then it is not difficult to see that 
\begin{equation}
	id_w(G) = id_{w'}(H).
\end{equation}
In other words, to solve the problem for $G$ we first solve the problem for
the subgraph $G[M]$, construct a new weighted graph $H$, and solve the problem for the graph $H$.

\subsubsection{Antineighborhood decomposition}
\label{subsec:anti}

One of the simplest branching algorithms for the maximum weight independent set problem 
is based  on the following obvious fact.  For any graph $G=(V,E)$ and any vertex $v \in V$,
$$
	is_w(G) = \max \{ is_w(G - N(v)), is_w(G - v) \},
$$
where $w$ is a weight function on the vertices of $G$, and $is_w(G)$ stands for the maximum weight of 
an independent set in $G$. We want to use a similar branching rule for the WID problem, i.e. 
\begin{equation}\label{eq:anti_WID}
	id_w(G) = \min \{ id_w(G - N(v)), id_w(G - v) \}.
\end{equation} 

However, formula (\ref{eq:anti_WID}) is not necessarily true, because
an independent dominating set in the graph $G - v$ is not necessarily dominating in the whole graph $G$. 
To overcome this difficulty, we introduce the following notion. 
\begin{definition}
A vertex $v$ is {\em permissible} if formula (\ref{eq:anti_WID}) is valid for $v$
\end{definition}

An obvious sufficient condition for a vertex to be permissible can be stated as follows:
if every independent dominating set in $G - v$ contains at least one neighbour of $v$, then $v$ is permissible.

Applying (\ref{eq:anti_WID}) to a permissible vertex $v$ of $G$, we reduce the problem from $G$ to two subgraphs $G - v$ and $G - N(v)$. 
Such a branching procedure results in a decision tree. 
In general, this approach does not provide a polynomial-time solution, since the decision tree may have exponentially many nodes (subproblems).
However, under some conditions this procedure may lead to a polynomial-time algorithm. In particular, this is true for graphs in hereditary classes 
possessing the following property. 

\begin{definition}\label{def:2}
A graph class ${\cal G}$ has the {\em antineighborhood property} 
if there is a subclass ${\cal F} \subseteq {\cal G}$, and polynomial algorithms $P, Q$ and $R$, such that 
\begin{enumerate}
	\item[(i)] Given a graph $G$ the algorithm $P$ decides whether $G$ belongs to ${\cal F}$ or not;
	\item[(ii)] $Q$ finds a permissible vertex $v$ in any input graph $G \in {\cal G} \setminus {\cal F}$ 
	such that the graph $G-N(v)$ induced by the antineighborhood of $v$ belongs to ${\cal F}$; 
	we call $v$ a {\em good vertex}; 	
	\item[(iii)] $R$ solves the WID problem for (every induced subgraph of) any input graph from ${\cal F}$.
\end{enumerate}
\end{definition}

Directly from the definition we derive the following conclusion.

\begin{theorem}\label{theo: anti}
Let ${\cal G}$ be a hereditary class possessing the antineighborhood property.
Then WID can be solved in polynomial time for graphs  in ${\cal G}$.
\end{theorem}

\subsubsection{Decomposition scheme}
\label{sec:had}

Let ${\cal G}$ be a hereditary class such that the class ${\cal G}_p$ of prime graphs in ${\cal G}$ 
has the antineighborhood property. 
We define the decomposition procedure by describing the corresponding decomposition tree $T(G)$ for 
a graph $G=(V,E) \in {\cal G}$. In the description, we use notions and notations introduced in Definition~\ref{def:2}.

\begin{enumerate}
	 \item If $G$ belongs to ${\cal F}$, then the node of $T(G)$ corresponding to $G$ is a leaf.

	\item If $G \not\in {\cal F}$ and $G$ has a homogeneous set $M$,
	then $G$ is decomposed into subgraphs $G_1 = G[M]$ and $G_2 = G[(V \setminus M) \cup \{m\}]$
	for some vertex $m$ in $M$. 
	The node of $T(G)$ corresponding to $G$ is called a \textit{homogeneous node}, and it has 
	two children corresponding to $G_1$ and $G_2$. These children are in turn the roots of subtrees
	representing possible decompositions of $G_1$ and $G_2$.

	\item If $G \not\in {\cal F}$ and $G$ has no homogeneous set, then $G$ is prime and by the
	 antineighborhood property of ${\cal G}_p$ there exists a good vertex $v \in V$. 
	 Then $G$ is decomposed into subgraphs $G_1 = G - N(v)$ and $G_2 = G - v$. 
	 The node of $T(G)$ corresponding to $G$ 
	 is called an \textit{antineighborhood node}, and it has two children corresponding to 
	 $G_1$ and $G_2$.
	 The graph $G_1$ belongs to ${\cal F}$ and the node corresponding to $G_1$ is a leaf. The node 
	 corresponding to $G_2$ is the root of a subtree representing a possible decomposition of $G_2$.

\end{enumerate}

\begin{lemma}\label{tree}
	Let $G$ be an $n$-vertex graph in ${\cal G}$. Then the tree $T(G)$ contains $O(n^2)$ nodes.
\end{lemma}
\begin{proof}
Since $T(G)$ is a binary tree, it is sufficient to show that the number of internal nodes is
$O(n^2)$. To this end, we prove that the  internal nodes of $T(G)$ can be labeled by 
pairwise different pairs $(a,b)$, where $a,b \in V(G)$.

Let $G' = (V',E')$ be an induced subgraph of $G$ that corresponds to
an internal node $X$ of $T(G)$.
If $X$ is a homogeneous node, then $G'$ is decomposed into
subgraphs $G_1 = G'[M]$ and $G_2 = G'[(V' \setminus M) \cup \{m\}]$, where $M \subset V'$ is a homogeneous
set of $G'$ and $m$ is a vertex in $M$. In this case, we label $X$ with $(a,b)$, where 
$a \in M \setminus \{m\}$ and $b \in V' \setminus M$.
If $X$ is an antineighborhood node, then $G'$ is decomposed into subgraphs 
$G_1 = G' - N(v)$ and $G_2 = G' - v$, where $v$ is a good vertex of $G'$. In this case, $X$ is labeled
with $(v,b)$, where $b \in N(v)$.

Suppose, to the contrary, that there are two internal nodes $A$ and $B$ in $T(G)$ with the
same label $(a,b)$. By construction, this means that $a,b$ are vertices of both $G_A$ and $G_B$, 
the subgraphs of $G$ corresponding to the nodes $A$ and $B$, respectively.
Assume first that $B$ is a descendant of $A$. The choice of the labels implies that
regardless of the type of node $A$ (homogeneous or antineighborhood), the label of 
$A$ has at least one vertex that is not a  vertex of $G_B$, a contradiction.
Now, assume that neither $A$ is a descendant of $B$ nor $B$ is
a descendant of $A$. Let $X$ be the lowest common ancestor of
$A$ and $B$ in $T(G)$. 
If $X$ is a homogeneous node, then $G_A$ and $G_B$ can have at most one vertex in 
common, and thus $A$ and $B$ cannot have the same label. 
If $X$ is an antineighborhood node, then one of its children is a leaf, contradicting to the
assumption that both $A$ and $B$ are internal nodes.  
\end{proof}

\begin{lemma}\label{lem:construct}
	Let $G$ be an $n$-vertex graph in ${\cal G}$. If time complexities of the algorithms $P$ and $Q$ are
	$O(n^p)$ and $O(n^q)$, respectively, then $T(G)$ can be constructed in time 
	$O(n^{2 + \max\{ 2, p, q \})})$.
\end{lemma}
\begin{proof}
	The time needed to construct $T(G)$ is the sum of times required to identify types of nodes of $T(G)$
	and to decompose graphs corresponding to internal nodes of $T(G)$. To determine the type
	of a given node $X$ of $T(G)$, we first use the algorithm $P$ to establish whether the graph 
	$G_X$ corresponding to $X$ belongs to ${\cal F}$ or not. In the former case $X$ is a leaf node, in the
	latter case we further try to find in $G_X$ a homogeneous set, which can be performed
	in $O(n+m)$ time \cite{McCSpi1999}. If $G_X$ has a homogeneous set, then $X$ is 
	a homogeneous node and we decompose $G_X$ into the graphs induced by the vertices in and outside
	the homogeneous set, respectively. If $G_X$ does not have a homogeneous set, then $X$
	is an antineighborhood node, and the decomposition of $G_X$ is equivalent to finding a
	good vertex, which can be done by means of the algorithm $Q$. 
	Since there are $O(n^2)$ nodes in $T(G)$, the total time complexity for constructing 
	$T(G)$ is $O(n^{2 + \max\{ 2, p, q \}})$. 	
\end{proof}


Now we are ready to prove the main result of this section.

\begin{theorem}\label{theo:decomposition}
	If ${\cal G}$ is a hereditary class such that the class ${\cal G}_p$ of prime graphs in ${\cal G}$ 
	has the antineighborhood property, then the WID problem 
	can be solved in polynomial time for graphs in  ${\cal G}$.
\end{theorem}
\begin{proof}
	Let $G$ be an $n$-vertex graph in ${\cal G}$. To solve the WID problem for $G$, we construct 
	$T(G)$ and then traverse it bottom-up, deriving a solution for each node of $T(G)$ from the solutions
	corresponding to the children of that node. 
	
	The construction of $T(G)$ requires a polynomial time by Lemma~\ref{lem:construct}. 
	For the instances corresponding to leaf-nodes of $T(G)$, the problem can be solved in polynomial time 
	by the antineighborhood property. 
	According to the discussion in Sections~\ref{subsec:modular} and~\ref{subsec:anti}, the solution for 
	an instance corresponding to an internal node can be derived from the solutions of its children 
	in polynomial time. 
	Finally, as there are $O(n^2)$ nodes in $T(G)$ (Lemma~\ref{tree}), the total running time to solve 
	the problem for $G$ is polynomial.
\end{proof}

\subsection{Application to $(P_5,\overline{P_5})$-free graphs}
\label{sec:P5}


In this section, we show that the WID problem can be solved efficiently for $(P_5,\overline{P_5})$-free 
graphs by means of the decomposition scheme described in Section~\ref{sec:had}.
To this end, we will prove that the class of prime $(P_5,\overline{P_5})$-free graphs has the antineighborhood property. 
We start with several auxiliary results. The first of them is simple and we omit its proof.  

\begin{observation}\label{obs:distAdj}
	Let $G=(V,E)$ be a graph, and let $W \subset V$ induce a connected
	subgraph in $G$. If a vertex $v \in V \setminus W$ distinguishes $W$, 
then $v$ distinguishes two	adjacent vertices of $W$.
\end{observation}

\begin{proposition}\label{st:distNonadj}
	Let $G=(V,E)$ be a prime graph. If a subset $W\subset V$ has at least two vertices and is not a clique, 
then there exists a vertex $v \in V \setminus W$ which distinguishes two non-adjacent vertices of $W$.
\end{proposition}
\begin{proof}
	Suppose, to the contrary, that none of the vertices in $V \setminus W$ distinguishes a pair of
	non-adjacent vertices in $W$. If $G[W]$ has more than one connected component, then it is
	easy to see that no vertex outside of $W$ distinguishes $W$. Hence,
	$W$ is a homogeneous set in $G$, which contradicts the primality of $G$. 

If $G[W]$ is connected, then $\overline{G[W]}$ has a connected component $C$ with at least two vertices, since
	$W$ is not a clique. Then, by our assumption and Observation~\ref{obs:distAdj}, no vertex outside of $W$ distinguishes $C$. 
Also, by the choice of $C$, no vertex of $W$ outside of $C$ distinguishes $C$. Therefore, $V(C)$
	is a homogeneous set in $G$. This contradiction completes the proof of the proposition.
\end{proof}

\begin{lemma}\label{lem:atom}
	If a $(P_5, \overline{P_5})$-free prime graph contains an induced copy of $2K_2$, then it has a 
	clique separator.
\end{lemma}
\begin{proof}
	Let $G=(V,E)$ be a $(P_5, \overline{P_5})$-free prime graph containing an induced copy of $2K_2$.
	Let $S \subseteq V$ be a minimal separator with the property that $G-S$ contains at least two non-trivial connected
components, i.e. connected components with at least two vertices. Such a separator necessarily exists, since $G$ contains an induced $2K_2$.

	It follows from the choice of $S$ that
	\begin{itemize}
		\item $G - S$ has $k \geq 2$ connected components $C_1, \ldots, C_k$;
		\item $r \geq 2$ of these components, say $C_1, \ldots, C_r$, have at least two vertices, and
		all the other components $C_{r+1}, \ldots, C_k$ are trivial;
		\item every vertex in $S$ has a neighbour in each of the non-trivial components
		$C_1, \ldots, C_r$ (since $S$ is minimal);
		\item for every $i \in \{ r+1, \ldots, k \}$, the unique vertex of the trivial component $C_i$
		has a neighbour in $S$ (since $G$ is connected).
	\end{itemize}
	
	In the remaining part of the proof, we show that $G$ has a clique separator.
	Let us denote 
	$U_i = V(C_i)$ for $i = 1, \ldots, k$. We first observe the following.

	\vskip1ex
	\textbf{Claim 1.} \textit{Any vertex in $S$ distinguishes at most one of the sets $U_1, \ldots, U_r$.}
	
	\vskip1ex
	\textit{Proof.} Assume $v \in S$ distinguishes $U_i$ and $U_j$ for distinct $i,j \in [r]$. Then by 
	Observation~\ref{obs:distAdj} $v$ distinguishes two adjacent vertices $a,b$ in $U_i$ and two adjacent
	vertices $c,d$ in $U_j$. But then $a,b,v,c,d$ induce a forbidden $P_5$.

	\vskip1ex	
	According to Claim 1, the set $S$ can be partitioned into subsets $S_0, S_1 \ldots, S_r$, 
	where the vertices of $S_0$
	dominate every member of $\{ U_1, \ldots, U_r \}$, and for each $i \in [r]$, the vertices
	of $S_i$ distinguish $U_i$ and dominate $U_j$ for all $j$ different from $i$.
	Moreover, for each $i \in [r]$ the set $S_i$ is non-empty, as the graph $G$ is prime. 
	Now we prove two more auxiliary claims.
	
	\vskip1ex
	\textbf{Claim 2.} \textit{For $0 \leq i < j \leq r$, every vertex in $S_i$ is adjacent to every vertex in $S_j$.}
	
	\vskip1ex
	\textit{Proof.} Assume that the claim is false, i.e. there exist two non-adjacent vertices $s_i \in S_i$
	and $s_j \in S_j$. By Observation \ref{obs:distAdj} there exist two adjacent vertices $a,b \in U_j$ that 
	are distinguished by $s_j$. But then $s_i, s_j, a, b$ and any vertex in $N(s_i) \cap U_i$ induce
	a forbidden $\overline{P_5}$, a contradiction.

	\vskip1ex
	\textbf{Claim 3.} \textit{For $i \in [r]$, no vertex in $U_i$ distinguishes two non-adjacent
	vertices in $S_i$.}
	
	\vskip1ex
	\textit{Proof.} Assume that there exists a pair of non-adjacent vertices $x,y \in S_i$ that are
	distinguished by a vertex $u_i \in U_i$. Let $j \in [r] \setminus \{ i \}$, and let $s_j \in S_j$
	and $u_j \in U_j \setminus N(s_j)$. Then, since $s_j$ dominates $S_i$, we have that $u_j, x, y, s_j, u_i$
	induce a forbidden $\overline{P_5}$, a contradiction.
	
	\vskip1ex
	We split further analysis into two cases.
	
	\textit{Case 1}: there is at least one trivial component in $G \setminus S$, i.e. $k > r$.
	For $i \in \{ r+1, \ldots, k \}$ we denote by $u_i$ the unique vertex of $U_i$.
	Let $U = \{ u_{r+1}, \ldots, u_k \}$ and let $u^*$ be a vertex in $U$ with a minimal (under inclusion)
	neighbourhood. We will show that $N(u^*)$ is a clique,
	and hence is a clique separator in $G$.
	By Claim 2, it suffices to show that $N(u^*) \cap S_i$ is a clique for each $i \in \{0, 1, \ldots, k\}$.
	Suppose that for some $i$ the set $N(u^*) \cap S_i$ is not a clique. 
	Then, by Proposition~\ref{st:distNonadj}, there are two nonadjacent vertices $x,y \in N(u^*) \cap S_i$
	distinguished by a vertex $z \in V \setminus (N(u^*) \cap S_i)$. It follows from Claims 2 and 3
	that either $z \in S_i \setminus N(u^*)$ or $z \in U$. If $z \in S_i \setminus N(u^*)$, then 
	$u^*, x, y, z,$ and any vertex in $U_j$, $j \in [r] \setminus \{ i \}$ induce a forbidden
	$\overline{P_5}$, a contradiction. 
	Hence, assume that none of the vertices in $S \setminus (N(u^*) \cap S_i)$ 
	distinguishes two nonadjacent vertices in $N(u^*) \cap S_i$.
	If $z \in U$, with $z$ being nonadjacent to $x$ and adjacent to $y$, then by the minimality 
	of $N(u^*)$
	there is a vertex $s \in N(z)$ that is not adjacent to $u^*$. Since $N(z) \subseteq S$, vertex $s$
	does not distinguish $x$ and $y$. But then $x, u^*, y, z, s$ induce either a $P_5$ (if $s$ is adjacent
	neither to $x$ nor to $y$) or a $\overline{P_5}$ (if $s$ is adjacent to both $x$ and $y$), a contradiction. 

	\vskip1ex

	\textit{Case 2}: there are no trivial components in $G \setminus S$, i.e. $k = r$.
	First, observe that $|S_0| \leq 1$, since $G$ is prime and no vertex outside of $S_0$ distinguishes $S_0$
	(which follows from the definition of $S_0$, Claim 2 and the fact that $k = r$).  Further,
	Claims 2 and 3 imply that for each $i \in [r]$ no vertex in $V \setminus S_i$ distinguishes two
	nonadjacent vertices in $S_i$. Therefore, applying  Proposition~\ref{st:distNonadj} we conclude  that 
	$S_i$ is a clique. Hence $S = \bigcup_{i=0}^{r} S_i$ is a clique separator in $G$.	
\end{proof}

\begin{lemma}\label{lem:perm}
	Let $G$ be a $(P_5, \overline{P_5})$-free prime graph containing an induced copy of $2K_2$.
	Then $G$ contains a permissible antisimplicial vertex.
\end{lemma}
\begin{proof}
	By Lemma \ref{lem:atom} graph $G$ has a clique separator, and therefore it also 
	has a minimal clique separator $S$.
	Let $C_1, \ldots, C_k$, $k \geq 2$, be connected components of $G-S$, and $U_i = V(C_i)$, 
	$i = 1, \ldots, k$.
	Since $S$ is a minimal separator, every vertex in $S$ has at least one neighbour in each of the sets
	$U_1, \ldots, U_k$.
	By Claim 1 in the proof of Lemma~\ref{lem:atom}, any vertex in $S$ distinguishes at most one of 
	the sets $U_1, \ldots, U_k$, and therefore, the set $S$ 
	partitions into subsets $S_0, S_1 \ldots, S_k$, where the vertices of $S_0$
	dominate every member of $\{ U_1, \ldots, U_k \}$, and for each $i \in [k]$ the vertices
	of $S_i$ distinguish $U_i$ and dominate $U_j$ for all $j$ different from $i$.
	
	If $S_0 \neq \emptyset$, then any vertex in $S_0$ is adjacent to all the other vertices in the graph,
	and therefore it is permissible and antisimplicial. Hence, without loss of generality, assume that 
	$S_0 = \emptyset$ and $S_1 \neq \emptyset$.
	
	Let $s$ be a vertex in $S_1$ with a maximal (under inclusion) neighbourhood in $U_1$.
	We will show that $s$ is antisimplicial and permissible.
	Suppose that the graph induced by the antineighbourhood of $s$ contains a
	connected component $C$ with at least two vertices. Since $G$ is prime, by Observation~\ref{obs:distAdj} it must 
	contain a vertex $p$ outside of $C$ distinguishing two adjacent vertices $q$ and $t$ in $C$. 
	Then $p$ does not belong to $N(s) \cap U_1$, since otherwise $q, t, p, s$ together with any
	vertex in $U_2$ would induce a $P_5$. Therefore, $p$ belongs to $S_1$.
	Since the set $N(s) \cap U_1$ is maximal, it contains a vertex $y$ nonadjacent to $p$. 
	But now $t, q, p, s, y$ induce either a $P_5$ or its complement, as $y$ does not distinguish 
	$q$ and $t$.
	This contradiction shows that every component in the graph induced by the antineighbourhood 
	of $s$ is trivial, i.e. $s$ is antisimplicial.
	
	Assume now that $s$ is not permissible, i.e. there exists an independent dominating set $I$ in 
	$G - s$ that does not contain a neighbour of $s$. Since $s$ dominates $U_2 \cup \ldots \cup U_k$,
	the set $I$ is a subset of $U_1 \setminus N(s)$. But then $I$ is not dominating, since no vertex of $U_2$ has a neighbour in $I$,
This contradiction completes the proof of the lemma.
\end{proof}

\begin{lemma}\label{lemm:anti}
	The class of prime $(P_5,\overline{P_5})$-free graphs has the antineighborhood
	property.
\end{lemma}
\begin{proof}
	Let ${\cal F}$ be the class of $(2K_2,\overline{P_5})$-free graphs (this is a subclass of 
	$(P_5,\overline{P_5})$-free graphs, since $2K_2$ is an induced subgraph of $P_5$). 
	Clearly, graphs in ${\cal F}$ can be recognized in polynomial time.
	Moreover, the WID problem can be solved in polynomial time for  graphs in ${\cal F}$, 
	because the problem is polynomially solvable on $2K_2$-free graphs (according to \cite{BY}, 
	these graphs have polynomially many maximal independent sets). 
	
	If a prime $(P_5,\overline{P_5})$-free graph $G=(V,E)$ does not belong to ${\cal F}$, then 
	by Lemma \ref{lem:perm} it contains a permissible vertex $v$ whose antineighbourhood is
	an independent set, and therefore, $G - N(v) \in {\cal F}$.
	It remains to check that a permissible antisimplicial vertex in $G$ can be found in polynomial time.
	It follows from the proof of Lemma~\ref{lem:perm} that in a minimal clique separator of $G$ 
	any vertex with a maximal neighbourhood is permissible and antisimplicial. A minimal
	clique separator in a graph can be found in polynomial time \cite{Whitesides1981}, and therefore
	the desired vertex can also be computed efficiently.
\end{proof}

Now the main result of the section follows from Theorem~\ref{theo:decomposition} and Lemma~\ref{lemm:anti}.

\begin{theorem}\label{thm:house}
	The WID problem is polynomial-time solvable in the class of $(P_5,\overline{P_5})$-free graphs.
\end{theorem}


\section{WID in \forbGraphs-free graphs}
\label{sec:new}

\begin{figure}[ht] 
     	\centering
     	\begin{tikzpicture}
			[scale=.6,auto=left]
	  		
	  		\node[w_vertex] (1) at (-1,2) {\footnotesize{$a_1$}};
	  		\node[w_vertex] (2) at (2,2) {\footnotesize{$a_2$}};
	  		\node[w_vertex] (3) at (2,-0.5) {\footnotesize{$a_3$}};
	  		\node[w_vertex] (4) at (-1,-0.5) {\footnotesize{$a_4$}};
	  		\node[w_vertex] (5) at (0.5,4) {\footnotesize{$a_5$}};
 	  		\node[w_vertex] (6) at (4.3,0.9) {$b$};
 	  		\node[w_vertex] (7) at (6.3,0.9) {$c$};
 	  		\node[w_vertex] (8) at (8.3,0.9) {$d$};
 
			\foreach \from/\to in {1/2,2/3,3/4,4/1,5/1,5/2,6/1,6/2,6/3,6/4,6/5,7/6,8/7}
	    		\draw (\from) -- (\to);
	\end{tikzpicture}
	    		
	\caption{The graph $\overline{P_5}^*$} 
	\label{fig:coP5star}
\end{figure}
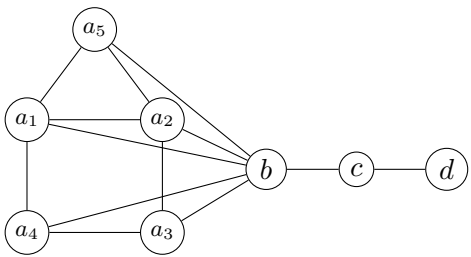

To solve the problem for  \forbGraphs-free graphs, let us introduce the following notation:
for an arbitrary graph $F$, we denote by $F^*$ the graph obtained from $F$ by adding three new vertices, say $b,c,d$,
such that $b$ dominates (adjacent to each vertex of) $F$, while $c$  is adjacent to $b$ and $d$ only (see Figure~\ref{fig:coP5star} for an illustration in the case $F=\overline{P}_5$).
The importance of this notation is due to the following result proved in \cite{LozMosPur2015}.
\begin{theorem}\label{th:P5Fstar}
	Let $F$ be any connected graph. If the WID problem can be solved in polynomial time for
	$(P_5,F)$-free graphs, then this problem can also be solved in polynomial time for $(P_5, F^*)$-free
	graphs.
\end{theorem}

This result together with Theorem~\ref{thm:house}  leads to the following conclusion. 
\begin{corollary}\label{cor:star}
The WID problem is polynomial-time solvable in the class of $(P_5,\compPfiveStar)$-free graphs.
\end{corollary}

To solve the problem for  \forbGraphs-free graphs, in this section we reduce it to  \threeForbGraphs-free graphs,  where the problem is solvable in polynomial time by Corollary~\ref{cor:star}.

\medskip
Let $G$ be a \forbGraphs-free graph containing a copy of $\compPfiveStar$ induced by vertices
$a_1, a_2, a_3, a_4, a_5, b, c, d$, as shown in Figure~\ref{fig:coP5star}. 

\noindent
Denote by $U$ the set of vertices in $G$ that have at
least one neighbour in $\{ a_1, a_2, a_3, a_4, a_5\}$, that is, $U = N(a_1) \cup \ldots \cup N(a_5)$.
In particular, $\{ a_1, a_2, a_3, a_4, a_5,b\}$ is a subset of $U$.
We assume that 
\begin{itemize}
\item[(**)] the copy of $\compPfiveStar$ in $G$ is chosen in such a way that $U$ has the minimum
number of elements. 
\end{itemize}
Now we prove several auxiliary results about the structure of $G$.

\begin{proposition}\label{prop:U2sep}
	If a vertex $x \in U$ has a neighbour $y$ outside of $U$, then $x$ is adjacent to each of the vertices
	$a_1, a_2, a_3, a_4$.
\end{proposition}
\begin{proof}
	Let $A = \{ a_1, a_2, a_3, a_4 \}$.
	Note that if $x$ is adjacent to $a_5$, then it must be adjacent to at least one vertex in $A$,
	since otherwise a forbidden $P_5$ arises.
	If $x$ is adjacent to exactly one or to exactly two adjacent vertices in $A$, then $\{ x,y \} \cup A$
	induces a subgraph containing a forbidden $P_5$.
	If $x$ is adjacent to exactly two non-adjacent vertices in $A$, say $a_1$ and $a_3$, then $x$
	must be adjacent to $a_5$, since otherwise $y,x,a_3,a_2,a_5$ induce a $P_5$. But this is
	impossible, since in this case $x,a_1,a_2,a_3,a_5$  induce a $\overline{P_3+P_2}$.
	Finally, if $x$ has exactly three neighbours in $A$, then $\{x\} \cup A$ induces a forbidden 
	$\overline{P_3+P_2}$. Therefore, $x$ must be adjacent to every vertex in $A$.
\end{proof}

\noindent
Taking into account Proposition~\ref{prop:U2sep}, we partition the set $U$ into three subsets as follows:
\begin{itemize}
	\item[$U_1$] consists of the vertices of $U$ that are adjacent to each of the vertices $a_1, a_2, a_3, a_4$,
	and have at least one neighbour outside of $U$;
	\item[$U_2$] consists of the vertices of $U$ that are adjacent to each of the vertices $a_1, a_2, a_3, a_4$,
	but have no neighbours outside of $U$;
	\item[$U_3$]$= U \setminus (U_1 \cup U_2)$.
\end{itemize}
Notice that $U_1$ is non-empty as it contains $b$. Also $\{ a_1, a_2, a_3, a_4, a_5 \} \subseteq U_3$, 
and no vertex in $U_3$ has a neighbour outside of $U$.

\begin{proposition}\label{prop:U2clique}
	$U_1$ is a clique in $G$.
\end{proposition}
\begin{proof}
	Suppose to the contrary that $U_1$ contains two non-adjacent vertices $x_1$ and $x_2$. Also, let $y_1$ and $y_2$ be neighbours of $x_1$ and $x_2$ outside of $U$,
	respectively. Vertex $y_1$ is not adjacent to $x_2$, since otherwise $x_1,x_2,a_1,a_2,y_1$
	induce a $\compPthreePtwo$. Similarly, $y_2$ is not adjacent to $x_1$. Hence $y_1 \neq y_2$,
	and therefore, to avoid a copy of $P_5$ induced by $y_1,x_1,a_1,x_2,y_2$, vertices $y_1$ and
	$y_2$ must be adjacent. For the same reason, $a_5$ should be adjacent to both $x_1$ and $x_2$.
	But then $x_1,x_2,a_3,a_4,a_5$ induce a copy of the forbidden $\compPthreePtwo$, a contradiction.
\end{proof}

\begin{proposition}\label{prop:coP5starFree}
	The graph $G[U_2 \cup U_3]$ is $\compPfiveStar$-free.
\end{proposition}
\begin{proof}
	Suppose to the contrary that $G[U_2 \cup U_3]$ contains vertices $a_1',a_2',a_3',a_4',a_5',b',c',d'$
	inducing a $\compPfiveStar$ (similarly to Figure~\ref{fig:coP5star}).
	Since no vertex in $U_2 \cup U_3$ has a neighbour outside of $U$ in $G$, and $c',d'$ are not adjacent
	to any of the vertices $a_1',a_2',a_3',a_4',a_5'$, we conclude that 
	$|N(a_1') \cup \ldots \cup N(a_5')| \leq |U|-2$, which contradicts the minimality of $|U|$.
\end{proof}

\medskip
Now we describe a reduction from the graph $G$ with a weight function $w$ to a graph $G'$
with a weight function $w'$, where $|V(G')| \leq |V(G)|-4$, $G'$ is \forbGraphs-free, and
$id_w(G) = id_{w'}(G')$.
First, we define $G'$ as the graph obtained from $G$ by
\begin{enumerate}
	\item removing the vertices of $U_3$;
	\item adding edges between any two non-adjacent vertices in $U_1 \cup U_2$;
	\item adding a new vertex $u$ adjacent to every vertex in $U_1 \cup U_2$.
\end{enumerate}

Clearly, $|V(G')| \leq |V(G)|-4$, as the set $U_3$ of the removed vertices contains at least 5 elements
and we add exactly one new vertex $u$. 
In the next proposition, we show that the above reduction does not produce any of the forbidden subgraphs.

\begin{proposition}\label{prop:GprimeForb}
	The graph $G'$ is \forbGraphs-free.
\end{proposition}
\begin{proof}
	Note that the graph $G' - (U_2 \cup \{u\})$ is isomorphic to $G - (U_2 \cup U_3)$, and therefore it contains
	no $P_5$ or $\compPthreePtwo$ as an induced subgraph. 
	Hence, if $G'$ contains a forbidden subgraph, then at least one of the vertices of this subgraph
	should lie in $U_2 \cup \{ u \}$.
	
	By construction of $G'$ and the definition of $U_2$, the set $U_2 \cup \{ u \}$ is a clique, and every vertex 
	in this set is simplicial in $G'$.
	Therefore, no vertex of $U_2 \cup \{ u \}$ can be a part of an induced copy of $\compPthreePtwo$.
	Also, $U_2 \cup \{ u \}$ can contain at most one vertex of an induced copy of $P_5$,
	and if $U_2 \cup \{ u \}$ contains such a vertex, it must be a degree-one vertex of the $P_5$. 
	Suppose to the contrary that $G'$ contains a copy of $P_5$ induced by $v_1,v_2,v_3,v_4,v_5$
	with $v_1 \in U_2 \cup \{ u \}$ and $\{ v_2,v_3,v_4,v_5 \} \subseteq V(G') \setminus (U_2 \cup \{ u \})$.
	But then $a_1,v_2,v_3,v_4,v_5$ induce a forbidden $P_5$ in $G$, a contradiction.
\end{proof}

\noindent
Now we define a weight function $w'$ on the vertex set of $G'$ as follows:
\begin{enumerate}
	\item $w'(x) = w(x)$, for every $x \in V(G') \setminus (\{ u \} \cup U_1 \cup U_2)$;
	\item $w'(u) = id_w(G[U_3])$;
	\item $w'(x) = w(x) + id_w(G[U \setminus N[x]])$, for every $x \in U_1$;
	\item $w'(x) = w(x) + id_w(G[U \setminus (U_1 \cup N[x])])$, for every $x \in U_2$.
\end{enumerate}

\begin{lemma}\label{lem:polyGprime}
	Given a weighted graph $(G,w)$, the weighted graph $(G',w')$ can be  constructed in polynomial time.
\end{lemma}
\begin{proof}
	To construct $G'$ we need to find in $G$ an induced copy of $\compPfiveStar$ that minimizes $|U|$. 
Clearly, this can be done in polynomial time.
	
	To show that $w'$ can be computed in polynomial time we observe that each of the graphs
	$G[U_3]$, 
	$G[U \setminus (U_1 \cup N[x])]$ for $x \in U_2$, and
	$G[U \setminus N[x]]$ for $x \in U_1$ is an induced subgraph of $G[U_2 \cup U_3]$.
	This observation together with Proposition~\ref{prop:coP5starFree} and Corollary~\ref{cor:star}
	imply the desired conclusion and finish the proof of the lemma.
\end{proof}

Now let us show that $id_w(G) = id_w'(G)$. 
For this, we will need two auxiliary propositions.

\begin{proposition}\label{prop:U3}
	Any independent dominating set in $G[U_3]$ dominates $U_1 \cup U_2$.
\end{proposition}
\begin{proof}
	Let $A = \{ a_1, a_2, a_3, a_4 \}$, and
	let $I$ be an independent dominating set in $G[U_3]$.
	If $I$ contains at least one of the vertices from $A$, then $I$ dominates $U_1 \cup U_2$,
	so we assume that $I \subseteq U_3 \setminus A$. Note that a vertex $x \in U_3 \setminus A$
	has at most two neighbours in $A$. Indeed, $x$ cannot have four neighbours by the definition of $U_3$,
	and it cannot have three neighbours, since otherwise $\{x\} \cup A$ induces a forbidden
	$\compPthreePtwo$.
	Now, if $I$ contains a vertex $x \in U_3 \setminus A$ that is adjacent  to
	$a_1$ and $a_3$, then $I$ dominates $U_1 \cup U_2$, since otherwise $x$ together with 
	$a_1, a_2, a_3$ and a non-neighbour of $x$ in $U_1 \cup U_2$ induce a forbidden 
	$\compPthreePtwo$.
	
	Assume that $I$ contains none of the above vertices. Then there exist vertices 
	$x,y \in I$ such that $x$ is adjacent to $a_1$ and non-adjacent to $a_3$, and $y$
	is adjacent to $a_3$ and non-adjacent to $a_1$. If $I$ does not dominate $U_1 \cup U_2$,
	then there exists a vertex $z \in U_1 \cup U_2$ that is adjacent neither to $x$ nor to $y$.
	But then $x,a_1, z, a_3, y$  induce a forbidden $P_5$.
 \end{proof}

\begin{proposition}\label{prop:U1}
	For every vertex $x \in U_2$, any independent dominating set in the graph $G - U$ dominates
	$U_1 \setminus N(x)$.
\end{proposition}
\begin{proof}
	Suppose to the contrary that there exists an independent dominating set $I$ in the graph $G-U$
	that does not dominate a vertex $y \in U_1 \setminus N(x)$.
	By the definition of $U_1$, vertex $y$ has a  neighbour $z$ in $V(G) \setminus U$.
	Since $I$ is dominating in $G-U$, there exists a vertex $v \in I$ that is adjacent to $z$.
	But then $v,z,y,a_1,x$ induce a forbidden $P_5$, a contradiction.
\end{proof}

\begin{lemma}\label{lem:weight}
	For any weighted graph $(G,w)$, we have $id_w(G) = id_{w'}(G')$.
\end{lemma}
\begin{proof}
	First, we show that $id_w(G) \geq id_{w'}(G')$.
	Let $I$ be an independent dominating set of the minimum weight in $G$. We distinguish between 
	the following three cases:
	\begin{enumerate}
		\item $I \cap U_1 \neq \emptyset$. \\
		By Propositions~\ref{prop:U2sep} and \ref{prop:U2clique}, the set $U_1$ is a clique separating
		$V(G) \setminus U$ from $U \setminus U_1$. Therefore, $I$ has only one element in $U_1$,
		say $x$, and:
		$$
			id_w(G) = w(x) + id_w(G[U \setminus N[x]]) + id_w(G - (U \cup N[x])).
		$$
		Consequently
		$$
			id_w(G) = w'(x) + id_{w'}(G' - N[x]) \geq id_{w'}(G').
		$$
		
		\item $I \cap U_1 = \emptyset$ and $I \cap U_2 \neq \emptyset$. \\
		Let $x \in I \cap U_2$. Then using Proposition~\ref{prop:U1}
		$$
			id_w(G) = w(x) + id_w(G[U \setminus (U_1 \cup N[x])]) + id_w(G - U) =
			w'(x) + id_{w'}(G' - N[x]) \geq id_{w'}(G').
		$$

		\item $I \cap (U_2 \cup U_1) = \emptyset$. \\
		In this case, taking into account Proposition \ref{prop:U3}, we conclude that
		$$
			id_w(G) = id_w(G[U_3]) + id_w(G - U) = w'(u) + id_{w'}(G' - N[u]) \geq id_{w'}(G').
		$$
	\end{enumerate}
	
	Let us now prove the reverse inequality $id_w(G) \leq id_{w'}(G')$.
	Let $I$ be an independent dominating set of the minimum weight in $G'$. Since $u$ does not have
	neighbours outside of $U_1 \cup U_2$, and $\{ u \} \cup U_1 \cup U_2$ is a clique in $G'$, the set $I$
	has exactly one element in $\{ u \} \cup U_1 \cup U_2$, which we denote by $x$.
	Similarly to the first part of the proof, we consider three cases:
	\begin{enumerate}
		\item $x \in U_1$. \\
		In this case
		$$
			id_{w'}(G') = w'(x) + id_{w'}(G' - N[x]) = w(x) + id_{w}(G[U \setminus N[x]]) + 
			id_w(G - (U \cup N[x])) \geq id_w(G).
		$$
	
		\item $x \in U_2$.\\
		In this case, by Proposition~\ref{prop:U1},
		$$
			id_{w'}(G') = w'(x) + id_{w'}(G'-N[x]) =
			w(x) + id_{w}(G[U \setminus (U_1 \cup N[x]) ]) + id_w(G-U) \geq id_w(G).
		$$
		
		\item $x = u$.\\
		In this case, by Proposition~\ref{prop:U3},
		$$
			id_{w'}(G') = w'(x) + id_{w'}(G'-N[x]) = id_w(G[U_3]) + id_w(G-U) \geq id_w(G).
		$$
	\end{enumerate}
\end{proof}

\noindent
Now we are ready to prove the main result of this section.

\begin{theorem}
	The WID problem is solvable in polynomial time for \forbGraphs-free graphs.
\end{theorem}
\begin{proof}
	Let $(G,w)$ be an $n$-vertex \forbGraphs-free weighted graph.
	If $G$ contains an induced copy of $\compPfiveStar$, then by Proposition~\ref{prop:GprimeForb}, and
	Lemmas~\ref{lem:polyGprime} and~\ref{lem:weight}, the graph $(G,w)$ can be transformed in polynomial
	time into a \forbGraphs-free weighted graph $(G',w')$ with at most $n-4$ vertices such that 
	$id_w(G) = id_{w'}(G')$.
	Repeating this procedure at most $\lfloor n/4 \rfloor$ times we obtain a \threeForbGraphs-free weighted
	graph $(H,\sigma)$ such that $id_w(G) = id_{\sigma}(H)$.
	By Corollary~\ref{cor:star} the WID problem for $(H,\sigma)$ can be solved in polynomial time.
	Finally, it is not difficult to see that a polynomial-time procedure computing $id_w(G)$ can be easily transformed into 
a polynomial-time algorithm finding an independent dominating set of weight $id_w(G)$.
\end{proof}

\section{Concluding remarks and open problems}

In this paper, we proved that \textsc{weighted independent domination}  can be solved
in polynomial time for $(P_5,\overline{P}_5)$-free graphs and  \forbGraphs-free graphs.  
A natural question to ask is whether these results can be extended to a class defined by one forbidden induced subgraph.

From the results in \cite{BL03} it follows that in the case of one forbidden induced subgraph $H$ the problem is solvable 
in polynomial time {\it only if} $H$ is a linear forest, i.e. a graph every connected component of which is a path. 
On the other hand, it is known that this necessary condition is not sufficient, since {\sc independent domination}
is NP-hard in the class of $2P_3$-free graphs. This follows from the fact that all sat-graphs are  $2P_3$-free \cite{Zverovich06}.

In the case of a {\it disconnected} forbidden graph $H$, polynomial-time algorithms to solve {\sc weighted independent domination} are known only for $mP_2$-free graphs for any fixed value of $m$.
This follows from a polynomial bound on the number of maximal independent sets in these graphs \cite{BY}.
The unweighted version of the problem can also be solved for $P_2+P_3$-free graphs \cite{LozMosPur2015}. 
However, for weighted graphs in this class the complexity status of the problem is unknown.

\begin{problem}
Determine the complexity status of {\sc weighted independent domination} in the class of $P_2+P_3$-free graphs. 
\end{problem}

In the case of a {\it connected} forbidden graph $H$, i.e. in the case when $H=P_k$, the complexity status is known for $k\ge 7$
(as $P_7$ contains a $2P_3$) and for $k\le 4$ (as  $P_4$-free graphs are precisely the cographs). Therefore, 
the only open cases are $P_5$-free and $P_6$-free graphs. 
As we mentioned in the introduction, the related problem of finding a maximum weight independent set (WIS) has been recently solved for $P_5$-free graphs \cite{P5}. 
This result makes the class of $P_5$-free graphs of particular interest for {\sc weighted independent domination}
and we formally state it as an open problem. 

\begin{problem}
Determine the complexity status of {\sc weighted independent domination} in the class of $P_5$-free graphs. 
\end{problem}

We also mentioned earlier that a polynomial-time solution for WIS
in a hereditary class $\cal X$ does not necessarily imply the same conclusion for WID in $\cal X$.
However, in the reverse direction such examples are not known. We believe that such examples do not exist and 
propose this idea as a conjecture. 

\begin{conjecture}
If WID admits a polynomial-time solution in a hereditary class $\cal X$, then so does WIS.
\end{conjecture}

\section*{Acknowledgements}

Vadim Lozin and Viktor Zamaraev acknowledge support of EPSRC, grant grant EP/L020408/1.
Dmitriy Malyshev was partially supported by Russian Foundation for Basic Research, grant No 16-31-60008-mol-a-dk; 
by RF President grant MK-4819.2016.1; by LATNA laboratory, National Research University Higher School of Economics.


\end{document}